# Performance Enhancement via XPM Suppression in a Linear all-PM NPE Mode-locked Fiber Oscillator

**MARVIN EDELMANN,**[1,2,*] **YI HUA,**[3,4] **MIKHAIL PERGAMENT,**[1] **AND FRANZ X. KÄRTNER**[1,2]

[1] *Center for Free-Electron Laser Science CFEL, Deutsches Elektronen-Synchrotron DESY, Notkestr. 85, 22607 Hamburg, Germany*
[2] *Department of Physics, Universität Hamburg, Jungiusstr. 9, 20355 Hamburg, Germany*
[3] *Deutsches Elektronen-Synchrotron DESY, Notkestr. 85, 22607 Hamburg, Germany*
[4] *European XFEL GmbH, Holzkoppel 4, 22869 Schenefeld, Germany*

*Corresponding author: marvin.edelmann@desy.de*

**We demonstrate strong performance enhancement of an all polarization-maintaining fiber oscillator mode-locked using NPE in a linear self-stabilized fiber interferometer via suppression of cross-phase modulation (XPM). Numerical simulations reveal that XPM significantly affects the saturable absorber dynamics resulting in distortions of mode-locked steady-states. In the experiment, we construct an oscillator with XPM suppression, employing an intra-cavity YVO₄ crystal, and compare its characteristics with a reference oscillator in standard configuration. It is shown, that XPM suppression not only lowers the mode-locking threshold by more than 40%, but further results in improved spectral pulse quality at the output ports and reduced nonlinear loss of the artificial saturable absorber.**

## 1. Introduction

Mode-locked fiber oscillators for the stable generation of ultrafast optical pulse trains are the driving source for manifold applications in science and technology, including frequency metrology [1], timing and synchronization in large scale research facilities [2] and high-resolution nonlinear biological imaging [3]. Over the past few decades, extensive research has therefore been directed at the development and improvement of high-performance oscillator setups. In particular oscillators mode-locked with *artificial* saturable absorbers (SA) based on the optical Kerr-effect, e.g., via implementation of nonlinear amplifying/optical loop mirrors (NALM/NOLM) [4-6] are known for their superior environmental stability, versatile mode-locking capabilities, and remarkable noise performance [7,8]. In contrast to NPE oscillators in ring configurations [9], NALM/NOLM oscillators leverage all-polarization maintaining (PM) fiber segments, significantly enhancing their long-term stability and resilience against environmental perturbations. They further offer a close to instantaneous response time for sub-100 fs pulse generation, and maintain a high threshold for optical degradation, setting them apart from *real* SAs such as semiconductor saturable absorber mirrors (SESAM) [10] or topological insulators [11,12].

In recent years, substantial research has been devoted to an alternative oscillator structure mode-locked via NPE in all-PM linear self-stabilized fiber interferometers (LSI) [13-16]. This concept, initially proposed by Fermann et al. in 1994 [17], provides a more streamlined cavity structure compared to NALM/NOLM mode-locked lasers and further offers new ways for adaptations, such as the recently demonstrated implementation of a coherent pulse divider for efficient energy scaling and noise reduction [18].

At first glance, the working principle of established NALM mode-locked oscillators, such as those in the Figure-of-9 configuration, appears closely related to that of LSI mode-locked oscillators. During each roundtrip within both cavities, the intra-cavity pulse undergoes a separation into two modes: two counterpropagating pulses in the NALM configuration and co-propagating polarization modes in the LSI configuration's PM-fiber segment. In both cases, these modes accumulate a differential nonlinear phase shift $\Delta\varphi_{nl}$, which is subsequently converted into self-amplitude modulation (SAM) through interaction with a polarizing element, such as a polarization beam-splitter (PBS).

However, the LSI oscillators introduce a unique element: the temporal overlap of both polarization modes in the fiber enables nonlinear phase distortions via cross-phase modulation (XPM) each roundtrip, a phenomenon not present in NALM lasers.

This study investigates the impact of XPM on the performance of LSI mode-locked fiber lasers and gives straight-forward guidelines for significant improvement. Numerical simulations reveal that XPM causes significant modulations in the output spectra via distortions of the artificial SA mechanism. In the experiment, we construct an LSI mode-locked fiber oscillator with XPM-suppression using an intra-cavity birefringent YVO4 crystal and compare it to a reference oscillator in standard configuration. In agreement with the numerical results, the XPM-suppression enables enhanced laser performance with a significant reduction in spectral modulations at the laser output ports, a reduced mode-locking threshold by more than 40% and a reduced nonlinear loss of the artificial saturable absorber mechanism.

## 2. Experimental Setup and Working Principle

The experimental setup of the enhanced/reference LSI mode-locked fiber oscillator is illustrated in Fig.1. Its all-PM fiber segment includes 0.7 m highly Ytterbium-doped fiber (YDF, CorActive Yb401-PM), optically pumped with a 1 W laser diode at 976 nm that is coupled to the YDF through a wavelength-division multiplexer (WDM). The free-space arm at the side of collimator C1 contains the cavity end-mirror followed by a 1000 lines/mm transmission grating pair (LightSmyth 1040-Series) for dispersion-management, a polarization beam-splitter (PBS) and the non-reciprocal phase-bias consisting of a Faraday-rotator (FR, 45° single-pass), a half-wave-plate (HWP) and an eight-wave plate (EWP) with tunable rotation angles $\theta_H$ and $\theta_E$, respectively. The quarter-wave plate (QWP) serves as tunable output coupler to monitor the intra-cavity field at Port T of PBS1. An additional half-wave plate (HWP1) is implemented to align the slow axis of the PM-fiber to the transmission axis of PBS1.

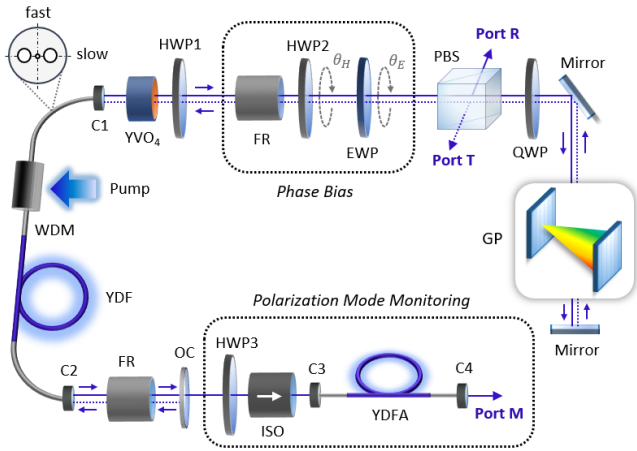

**Fig. 1:** Experimental setup of the enhanced LSI mode-locked fiber oscillator with intra-cavity YVO4 crystal and monitor port for monitoring the separated polarization modes. The reference setup is obtained by removing the YVO4-crystal. GP, grating pair; QWP, quarter-wave plate; PBS, polarization beam-splitter; EWP, eight-wave plate; HWP, half-wave plate; FR, Faraday-rotator; C, Collimator; WDM, wavelength-division multiplexer; YDF, Ytterbium-doped fiber; OC, output coupler; ISO, isolator; YDFA, Ytterbium-doped fiber amplifier.

Each roundtrip in the LSI cavity, the phase-bias rotation angles $\theta_E$ and $\theta_H$ generate orthogonal polarization modes parallel to fast and slow axis of the PM-fiber with fixed energy-splitting ratio and phase-offset. For an asymmetric energy-splitting ratio, both modes accumulate a nonlinear phase-difference $\Delta\varphi_{nl}$ via the optical Kerr-effect. Simultaneously, the fibers birefringence leads to a difference in linear phase-shift, causing a walk-off between both modes by ~1.25 ps/m (PM980-XP fiber at 1030 nm). To compensate this drift-off, an adapted Faraday-rotator mirror (FRM) is attached at the side of C2, consisting of a FR (45° single-pass) and an 80:20 output coupler. After double pass through the PM-fiber, the compensation of the linear phase difference ensures the overlap of both polarization modes in time with an accumulated $\Delta\varphi_{nl}$ that results in a nonlinear polarization rotation (NPR) of the recombined field. At the PBS, this NPR results in a separation of the intra-cavity field into a reflected (Port R) and transmitted (Port T) component, depending on the characteristic sinusoidal transmission function $T(\Delta\varphi_{nl})$. Shape and offset of $T(\Delta\varphi_{nl})$ is tunable via $\theta_E$ and $\theta_H$ and fully characterizes the artificial SA [20].

The slow walk-off between the polarization modes during nonlinear propagation in the PM-fiber enables phase-distortions via XPM. The setup can take the form of a reference setup (RS) with XPM and an enhanced setup (ES) with XOM-suppression. For the ES, a highly birefringent 20 mm long YVO4 crystal is included between C1 and HWP1 to pre-separate both modes in time suppress any nonlinear interaction via XPM in the fiber segment.

## 3. Numerical Simulations

To investigate the influence of XPM on mode-locked steady-states in LSI oscillators, numerical simulations are applied. To include XPM, the nonlinear propagation in the fiber segment is described via coupled nonlinear Schrödinger equations (NLSE) similar to Ref. [19]. The model further includes Kerr nonlinearity, chromatic dispersion up to the second order [20,21] and gain based on rate-equations which account for the double pass configuration [22]. The artificial SA is described via Jones formalism to derive the SA transmission $T(\Delta\varphi_{nl})$ similar to Ref. [7]. The simulation parameters are matched to the experimental setup in Fig.1, with $\theta_H = 52°$ and $\theta_E = 81°$ for optimum starting conditions with maximum positive slope at $T(\Delta\varphi_{nl} = 0)$. The QWP rotation angle is set to ensure 40% output coupling ratio to Port T. The free-space cavity loss, considering the insertion loss of the GP, coupling efficiency of C1/2 and the OC is lumped at the side of C1 and C2 with 70% and 40%, respectively.

For the reference oscillator (RS) in mode-locked steady-state, Fig.2 [a] shows the simulated evolution of the intensity-dependent $\Delta\varphi_{nl}$ influenced by XPM as function of the propagation distance in the RS fiber segment. Here, the position at 0 m corresponds to the position of the FRM. As shown, $\Delta\varphi_{nl}$ accumulates significant modulations during the nonlinear propagation, in particular during a recombination of the amplified polarization modes in backward propagation from the FRM in the direction to C1 (Fig.1). The modulations of $\Delta\varphi_{nl}$ at the fiber output shown in Fig1 [b], which is directly related to the SA transmission via $T(\Delta\varphi_{nl})$, results in modulations of the output spectral densities emitted at Port T and R of the LSI oscillator (Inset Fig.2 [a]). Similar spectral distortions have been reported over a broad range of LSI oscillator cavity parameters and mode-locking regimes, in a variety of experimental studies; e.g., in Ref. [13-16].

In contrast, Fig.2 [c] shows the evolution of $\Delta\varphi_{nl}$ with identical cavity parameters but in the ES with XPM suppression via YVO4 crystal. In this case, an interaction of polarization modes via XPM is prevented by a pre-separation in time as illustrated. The XPM suppression strongly reduces the modulations in the overall accumulated $\Delta\varphi_{nl}$ (Fig.2 [d]) and therefore the distortions of the artificial SA transmission. As consequence, smooth output spectra at Port T and R (inset of Fig.2 [c]) can be obtained. The absence of spectral modulations indicates significant potential for overall performance enhancement of LSI oscillators via XPM suppression.

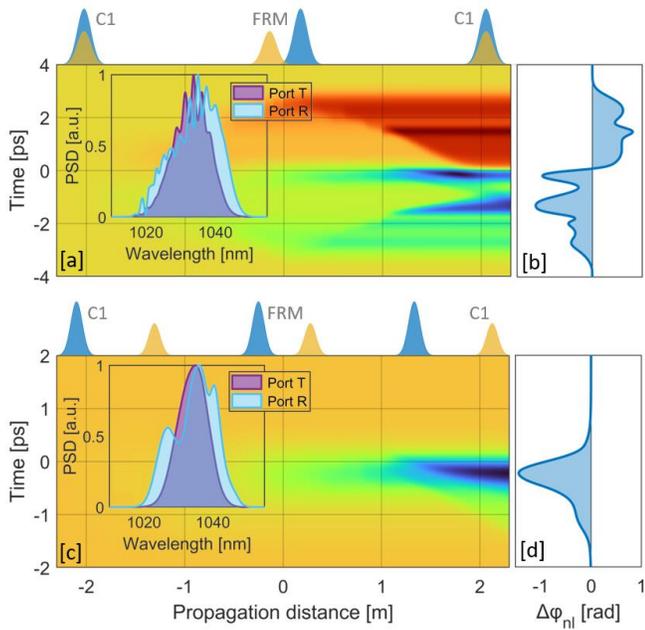

**Fig. 2:** [a]: Simulated evolution of $\Delta\varphi_{nl}(t)$ in the RS with XPM as function of the fiber segment propagation distance together with the position-dependent polarization mode delay. The inset shows the resulting output spectra at Port T (blue) and R (yellow). [b]: Corresponding total roundtrip $\Delta\varphi_{nl}(t)$. [c]: Evolution of $\Delta\varphi_{nl}(t)$ in the ES with XPM-suppression. Inset: Steady-state Port T (blue) and R (yellow) spectra. [d]: Total roundtrip $\Delta\varphi_{nl}(t)$.

## 4. Experimental Results and Discussion

To investigate the influence of XPM suppression experimentally, comparative measurements are conducted on the reference setup (RS) and enhanced (XPM-suppressed) setup (ES) with YVO$_4$ crystal. To ensure comparability, both setups are operated in a stretched-pulse mode-locking regime with $\sim -5*10^3\ fs^2$ net dispersion. For both RS and ES, the cw-laser threshold is measured at $\sim$60 mW, verifying a negligible influence of the YVO$_4$ crystal on the cavity alignment and *linear* loss.

To investigate the influence of XPM on the mode-locking threshold, Fig.3 [a] illustrates both oscillators slope efficiencies $dP_{out}/dP_p$ ($P_p$: pump power, $P_{out}$: Port T output power) together with the corresponding power transfer curve (inset). For this measurement, the pump power is step-wise increased in 25 mW steps up to a maximum of 1.5 W. The transition from cw to the energetically more efficient mode-locked state is clearly visible by a sharp increase in slope efficiency. Compared to the RS with a measured mode-locking threshold at 1.225 W, the avoidance of XPM in the ES results in a reduced threshold at 0.7 W, corresponding to $\sim$43% less required pump power. The reduced starting ability of the RS can be explained with XPM-induced distortions and a resulting increase in nonlinear loss of the artificial SA mechanism that is verified later on.

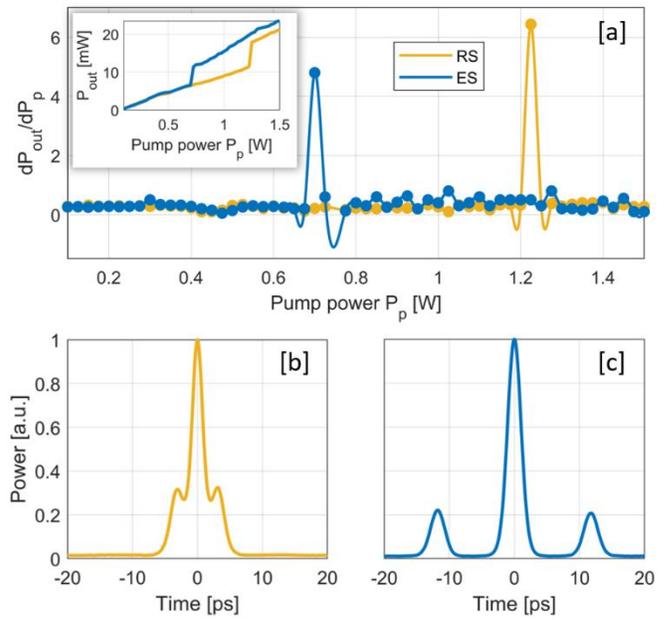

**Fig. 3:** [a]: Measured slope efficiency $dP_{out}/dP_p$ as function of the pump power $P_p$ for the RS (blue) and ES (yellow). The inset shows the corresponding power curve $P_{out}(P_p)$. [b]: AC trace of the separated polarization modes measured at Port M of the RS. [b]: Corresponding AC trace of the ES.

Once mode-locking is initiated, stable single-pulse operation can be obtained by reducing the pump power to $\sim$50 mW for the RS and 55 mW for the ES. Fig.3 [b] and [c] show AC traces of the separated polarization modes with maximum delay measured at Port M in single-pulse regime of the RS and ES, respectively. In the case of the RS, the measured $\sim$ 3 ps walk-off is entirely caused by the PM-fibers birefringence. For the ES, YVO$_4$ crystal with a birefringence of 0.208 at 1030 nm and 20 mm length causes a pre-separation of $\sim$16 ps. In the experiment, the fast axis of the YVO$_4$ is aligned parallel to the slow axis of the PM-fiber, resulting in the measured walk-off by $\sim$13 ps shown in Fig. 3 [c].

In LSI mode-locked oscillators, the output power at Port T is directly proportional to the intra-cavity power, while the Port R output power reflects the loss of the artificial SA. A comparison of their power ratio therefore yields information about the relative nonlinear loss introduced by the artificial SA mechanism, independent of the pump power. The measured output power for the RS at Port T and Port R is 0.6 mW and 2 mW corresponding to pulse energies of 16.7 pJ and 55.6 pJ, respectively. Conversely, the measured output power at the corresponding ES output ports are 1.4 mW and 0.9 mW with pulse energies of 38.9 pJ and 25 pJ, respectively. Notably, the power ratio between Port T and Port R is inverted between RS and ES. In the ES with XPM-suppression, the Port T output power exceeds the Port R output power by $\sim$ 64 %. In contrast for the RS, the output power at Port R exceeds that of Port T by more than 300%. Considering the verified identical linear cavity loss for ES and RS, the suppression of XPM in the ES therefore seems to improve not only the artificial SAs self-starting capability (Fig.3 [a]) but further its overall nonlinear transmission in mode-locked steady-state.

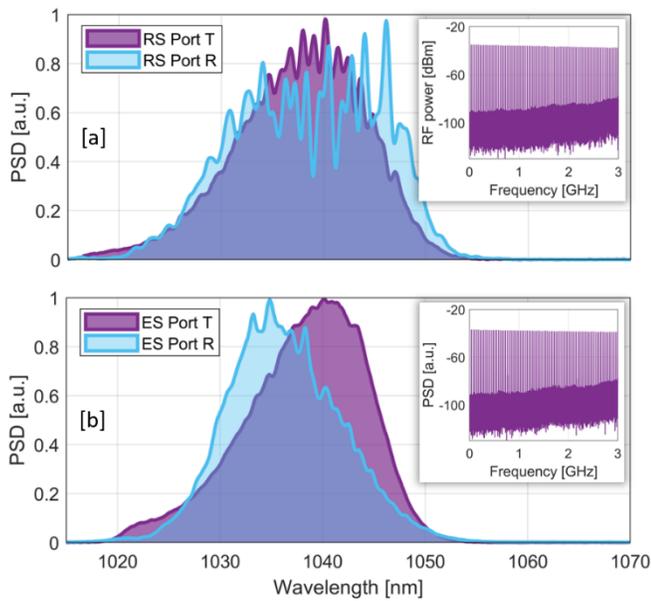

**Fig. 4:** [a]: Output spectra of the RS in stretched-pulse regime measured at Port T (blue) and Port R (orange) [b]: Spectra measured at Port T and R of the ES with XPM-suppression. The inset show the broadband RF spectra measured at Port T of the respective configuration.

To further investigate the characteristics of this energy transfer difference and also the general quality of the output pulses in both configurations, Fig. 4 [a] and [b] illustrate the Port T/R output spectra measured for the RS and ES, respectively. To verify the occurrence of periodic perturbations predicted by the simulations with high fidelity, the spectra are measured with an optical spectrum analyzer (ANDO AQ6315A) at a high spectral resolution of 0.02 nm. The insets in Fig.4 [a] and [b] show the broadband RF spectra measured at Port T of the ES and RS, respectively. Measured with a fast 9 GHz photodetector (ET-3500AF) and a RF spectrum analyzer (Agilent N9030A), the flat amplitudes of the higher harmonics are verified to ensure single-pulse operation for the optical spectrum measurements in both configurations.

As shown in Fig.4 [a], the measured RS output spectra at both Port T and R are strongly modulated with a full width at half maximum (FWHM) of 13.8 nm and 18.1 nm respectively. In comparison, the output spectra at Port T and R of the ES in Fig.4 [b] have a FWHM of 13.4 nm and 12.2 nm with significantly reduced spectral modulations. This result verifies the suppression of XPM as crucial precondition to obtain a high spectral output quality from LSI oscillators comparable to that of e.g., NALM/NOLM and NPE oscillators.

## 5. Conclusion

In conclusion, we demonstrated the significant potential of XPM suppression for performance enhancement of all-PM LSI mode-locked fiber oscillators. Numerical simulations are conducted to verify the influence of XPM as major contribution to distortions of the artificial saturable absorber mechanism. In the experiment, an oscillator with XPM-suppression via intra-cavity $YVO_4$ crystal is constructed and compared to a reference laser. In strong agreement with the numerical results, it is shown that the suppression of XPM results in a reduction of the mode-locking threshold by more than 45 %, a reduction in nonlinear loss of the artificial SA and a significant reduction in spectral modulations at the laser output ports. The demonstrated performance enhancement will greatly increase the range of applications for LSI mode-locked fiber lasers and enable real competition with high performance NALM/NOLM mode-locked oscillators.

**Funding.** Deutsche Forschungsgemeinschaft (KA 908-1MUJO), European Research Council (609920).

**Disclosures.** The authors declare no conflicts of interest.

**Data availability.** Data underlying the results presented in this paper are not publicly available at this time but may be obtained from the authors upon reasonable request.